\input{aipcheck}

\documentclass[
    ,final            
  ]
  {aipproc}

\newcommand{\aap}{{ A\&A~\/}}
\newcommand{\aaps}{{ A\&AS~\/}}
\newcommand{\aj}{{ AJ~\/}}
\newcommand{\apj}{{ ApJ~\/}}
 
\newcommand{\apjs}{{ ApJS~\/}}

\newcommand{\mnras}{{ MNRAS~\/}} 

\newcommand{\pasp}{{ PASP~\/}}
\newcommand{\nat}{{ Nature~\/}}

\layoutstyle{6x9}

\def\mjup{\mbox{$M_\textrm{\tiny Jup}$}}
\def\msun{\mbox{$M_\odot$}}

\def\um{\mbox{$\mu$m}}
\def\ga{\mathrel{\hbox{\rlap{\hbox{\lower4pt\hbox{$\sim$}}}{\raise2pt\hbox{$>$}}}}}\def\la{\mathrel{\hbox{\rlap{\hbox{\lower4pt\hbox{$\sim$}}}{\raise2pt\hbox{$<$}}}}}

\begin{document}

\title{Brown Dwarf Companions to White Dwarfs}


\author{Matt R. Burleigh}{ 
  address={Dept. of Physics and Astronomy, University of Leicester, Leicester, LE1 7RH, UK}
}

\author{Paul R. Steele}{
  address={Max-Planck-Institut f\"ur extraterrestrische Physik, Giessenbachstra\ss{}e, 85748, Garching, Germany}
}

\author{Paul D. Dobbie}{
  address={Australian Astronomical Observatory, PO Box 296, Epping, Sydney, NSW 1710, Australia}
}

\author{Jay Farihi}{ 
  address={Dept. of Physics and Astronomy, University of Leicester, Leicester, LE1 7RH, UK}
}

\author{Ralf Napiwotzki}{
  address={Centre for Astrophysics Research, University of Hertfordshire, Hatfield AL10 9AB, UK}
}

\author{Pierre F. L. Maxted}{
  address={Astrophysics Group, Keele University, Staffordshire, ST5 5BG, UK}
}

\author{Martin A. Barstow}{ 
  address={Dept. of Physics and Astronomy, University of Leicester, Leicester, LE1 7RH, UK}
}

\author{Richard F. Jameson}{ 
  address={Dept. of Physics and Astronomy, University of Leicester, Leicester, LE1 7RH, UK}
}

\author{Sarah L. Casewell}{ 
  address={Dept. of Physics and Astronomy, University of Leicester, Leicester, LE1 7RH, UK}
}




\author{Boris T. Gaensicke}{
  address={Department of Physics, University of Warwick, Coventry, CV4
  7AL, UK}
}

\author{Tom R. Marsh}{
  address={Department of Physics, University of Warwick, Coventry, CV4
  7AL, UK}
}

\begin{abstract}
Brown dwarf companions to white dwarfs are rare, but recent infra-red
surveys are slowly revealing examples. We present new observations of
the post-common envelope binary WD0137$-$349, which reveals the
effects of irradiation on the $\approx0.05\msun$ secondary, and new
observations of GD\,1400 which show that it too is a close, post-common
envelope system.  We also present the latest results in a
near-infrared photometric search for unresolved ultra-cool companions
and to white dwarfs with UKIDSS. Twenty five DA white
dwarfs were identified as having photometric 
excesses indicative of a low
mass companion, with 8-10 of these having a predicted mass in the
range associated with brown dwarfs. The results of this survey show
that the unresolved ($<2$'') brown dwarf companion fraction to DA
white dwarfs is $0.3\le f_{\rm WD+BD}\le1.3$\%.
\end{abstract}

\maketitle


\section{Introduction}

Large-scale near-infrared (NIR) surveys of white dwarfs (WDs) are
ideal for the searches and studies of substellar companions. A typical
WD is $10^3 - 10^4$ times fainter than its main sequence (MS)
progenitor, significantly reducing the brightness contrast problem
when searching for cool, low mass secondaries. In addition, the
spectral energy distributions of WDs (blue) and their low mass
companions (red) are markedly different, facilitating easy separation
of the components in broadband photometry and enabling straightforward
spectroscopic follow-up (e.g.\ \citealt{dobbie05}).

Searches for substellar companions to WDs allows for the investigation
of the known deficit of brown dwarf (BD) companions to MS stars 
\citep{mccarthy,grether}. Since the ages of WDs can be relatively well
constrained,  any BD companions can potentially 
be regarded as ``benchmarks'' for
testing evolutionary and atmospheric models \citep{pinfield06}.
The closest WD$+$BD binaries might
also represent either a channel for cataclysmic variable (CV) 
evolution or the end
state of CV evolution, in which the secondary has become highly
evolved through mass transfer \citep{patterson05}. Indeed, there are
a growing number of CVs with confirmed BD secondaries
(e.g.\ \citealt{littlefair1, littlefair2}). 
In close detached binaries, the BD is
expected to be irradiated by the WD's high UV flux, possibly leading
to substantial temperature differences between the ``day'' and
``night'' hemispheres. Such systems can provide laboratories for
testing models of irradiated "hot Jupiter" atmospheres
(e.g.~HD\,189733b; \citealt{knutson07}). However, detached BD
companions to WDs are rare. \cite{fbz05} calculate a binary
fraction of $f_{\rm WD+dL}<0.5$\% for L dwarfs.  

Radial velocity and
proper motion surveys, and searches for NIR excesses have so far found
only five confirmed examples: GD\,165 (DA$+$dL4, \citealt{becklin88}), GD\,1400
(DA$+$dL$6-7$; \citealt{gd1400,dobbie05}),
WD\,$0137-349$ (DA$+$dL8; \citealt{maxted06,wd0137b}),
PHL\,5038 (DA$+$dL8; \citealt{steele09}), and the first T dwarf $+$ WD
pair, LPSM\,$1459+0857$\,AB \citep{avril10}. 
GD\,165, PHL\,5038 and LPSM\,$1459+0857$\,AB 
are wide systems with projected separations of 120\,AU, 
55\,AU and $\sim 20,000$\,AU 
respectively, whereas GD\,1400 and WD\,0137$-$349 are in much
closer orbits with periods of $\approx10$~hours (see below)
and 116~mins respectively.

\section{New observations of the close binaries WD\,0137$-$349 and GD\,1400}

\begin{figure}[ht]
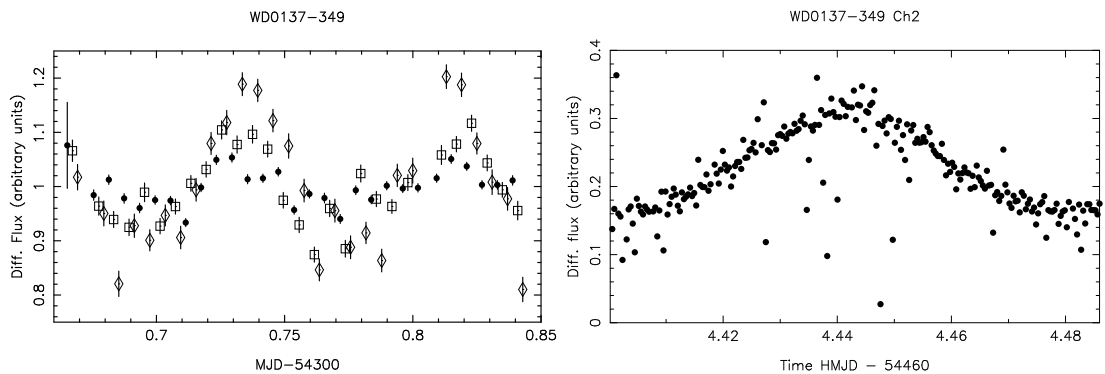

  \includegraphics[width=4.8cm,angle=-90]{mrburleigh_2_f1a.ps}
  \includegraphics[width=4.8cm,angle=-90]{mrburleigh_2_f1b.ps}
  \caption{Right: Near-IR light curves of WD0137$-$349 obtained with
 IRIS2 on the AAT. Filled circles: $J$-band, open squares: $H$-band,
 open diamonds: $K$-band. Left: Mid-IR, {\it Spitzer} $4.5\mu$m 
light curve.}
  \label{WD0137}
\end{figure}

\begin{figure}[ht]
\parbox{\textwidth}{%
\begin{center}
\includegraphics[height=.27\textheight, angle=0]{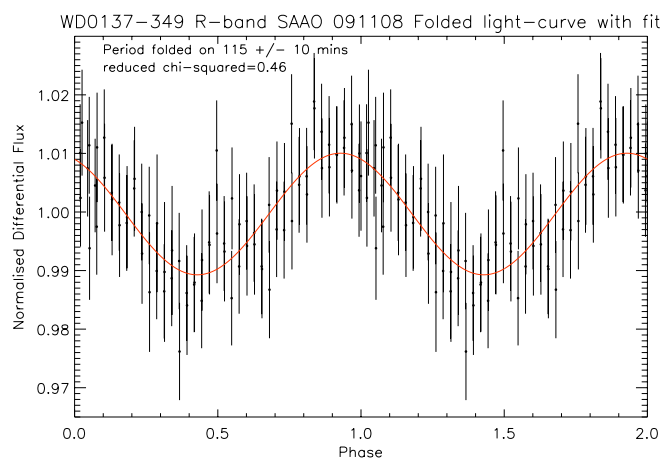}
\caption{Optical $R$-band light curve of WD0137$-$349 obtained with
 the SAAO 1.0m.}
\label{WD0137b}
\end{center}
}
\end{figure}


We have continued to observe WD\,0137$-$349 since its discovery by 
\cite{maxted06}. Multi-waveband light curves from the optical to the
mid-infrared show that the BD is being heated on the
hemisphere facing the WD, producing significant variability on the 
$116$~min orbital period, from
$\pm10$\% at $4.5\um$ to $\pm1$\% at $R$ (Figure \ref{WD0137}and 
Figure \ref{WD0137b}). 
The optical modulations are
a slightly surprising discovery, and we are modeling these light
curves to further investigate the effects of irradiation on the
substellar component's atmosphere.   


\begin{figure}[ht]
  \includegraphics[width=8cm,angle=-90]{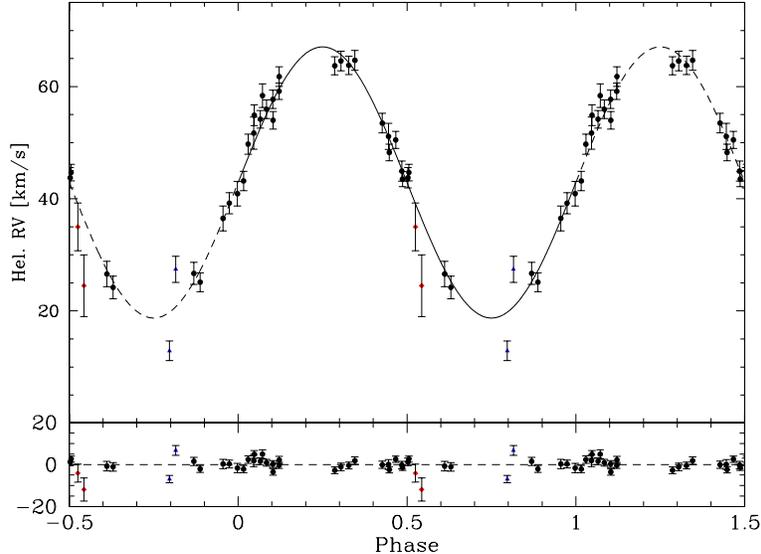}
  \caption{Radial velocity measurements of the white dwarf component of
  GD1400, obtained with UVES on the VLT, 
folded on the best-fit period of 9.98~hours.}
  \label{gd1400a}
\end{figure}


Radial velocity observations of
GD\,1400 obtained with UVES on the VLT 
 (Burleigh et al.\ 2011, in prep.) conclusively
show that it is a close, post-common envelope binary with an orbital
period $P_{\rm orb}=9.98$~hrs (Figure~\ref{gd1400a}). Although 
H~emission from the brown dwarf is tentatively detected 
at one phase (possibly due to a flare
rather than irradiation), Burleigh et al. are unable to
determine its radial velocity. From the mass function 
they place a lower limit on the BD mass $M_2 \approx 0.06\msun$, with
a lower limit on the WD mass $M_2 \approx 0.55\msun$. 

Near-IR photometry of GD\,1400, obtained sporadically across the
9.98~hr orbital period by Burleigh et al. (2011, in prep.), 
shows no variability. This is perhaps
unsurprising: the greater separation of the two components compared to
WD0137$-$349, and the cooler WD primary ($T_{\rm eff}\approx11,500$K
versus $T_{\rm eff}\approx16,000$K for WD0137$-$349) suggests
much weaker irradiation of the BD secondary. 

GD\,1400B is the second substellar
object that must have survived engulfment by its parent star's
atmosphere during one of the giant phases of stellar evolution, after 
WD\,$0137-349$. From
consideration of the white dwarf masses, it is likely that GD\,1400
underwent common envelope evolution on the asymptotic giant branch 
 and that the brown dwarf
originally orbited the main sequence progenitor star between $0.1$ and
$1$\,AU, while WD\,$0137-349$ went through a common envelope stage on
the red giant branch 
and had an original orbital separation $\la 0.1$\,AU. These
binaries are the direct descendants of 
intermediate-mass giant stars with brown dwarf companions in orbits
$a\sim$~few~AU.

The WD component of GD\,1400 is also a ZZ Ceti
pulsator. \cite{fontaine03} identified three significant frequency
components at 823.2s, 727.9s and 462.2s, the latter having the
strongest amplitude. 

\section{A survey for WD$+$BD binaries with UKIDSS}


New large infra-red sky surveys will reveal more examples of WD$+$BD
binaries, and allow us to better constrain their frequency. Here, we
present preliminary results from the UKIRT Infrared Deep Sky Survey
(UKIDSS, \citealt{ukidss1,ukidss2}). 
UKIDSS is the near-IR counterpart of the Sloan Digital Sky Survey
 and is several magnitudes deeper than the previous
2MASS survey, although unlike the latter
UKIDSS will cover only $\sim20\%$ of the sky. 
UKIDSS uses the 0.21\,deg$^2$ field of view 
Wide Field Camera  \citep{ukidss3} 
on the 3.8m UKIRT telescope on Mauna Kea, Hawaii. The
UKIDSS surveys began in May 2005 and are expected to be completed in 
2012. The survey of most relevance to this work is the Large Area
Survey (LAS), which aims to cover 
$\sim4000$\,deg$^2$ of the Northern Sky coincident with the SDSS. The
LAS makes observations in the $YJHK$ filters to a $5\sigma$
depth for point sources of $Y \approx 20.2$, $J \approx 19.6$, $H
\approx 18.9$ and $K \approx 18.2$ \citep{ukidss4}. 

\begin{figure}
  \includegraphics[width=4.5cm,angle=-90]{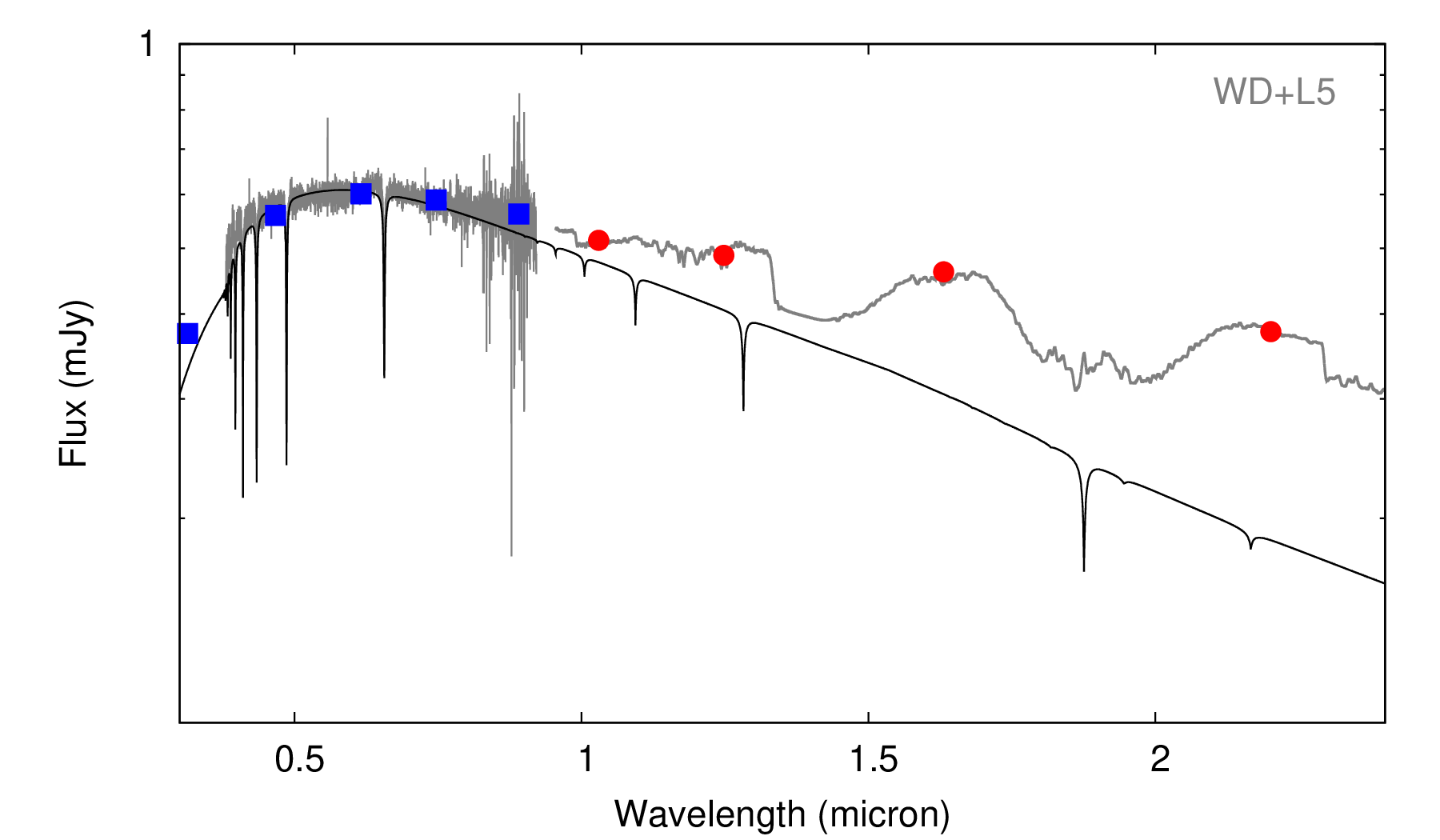}
  \includegraphics[width=4.5cm,angle=-90]{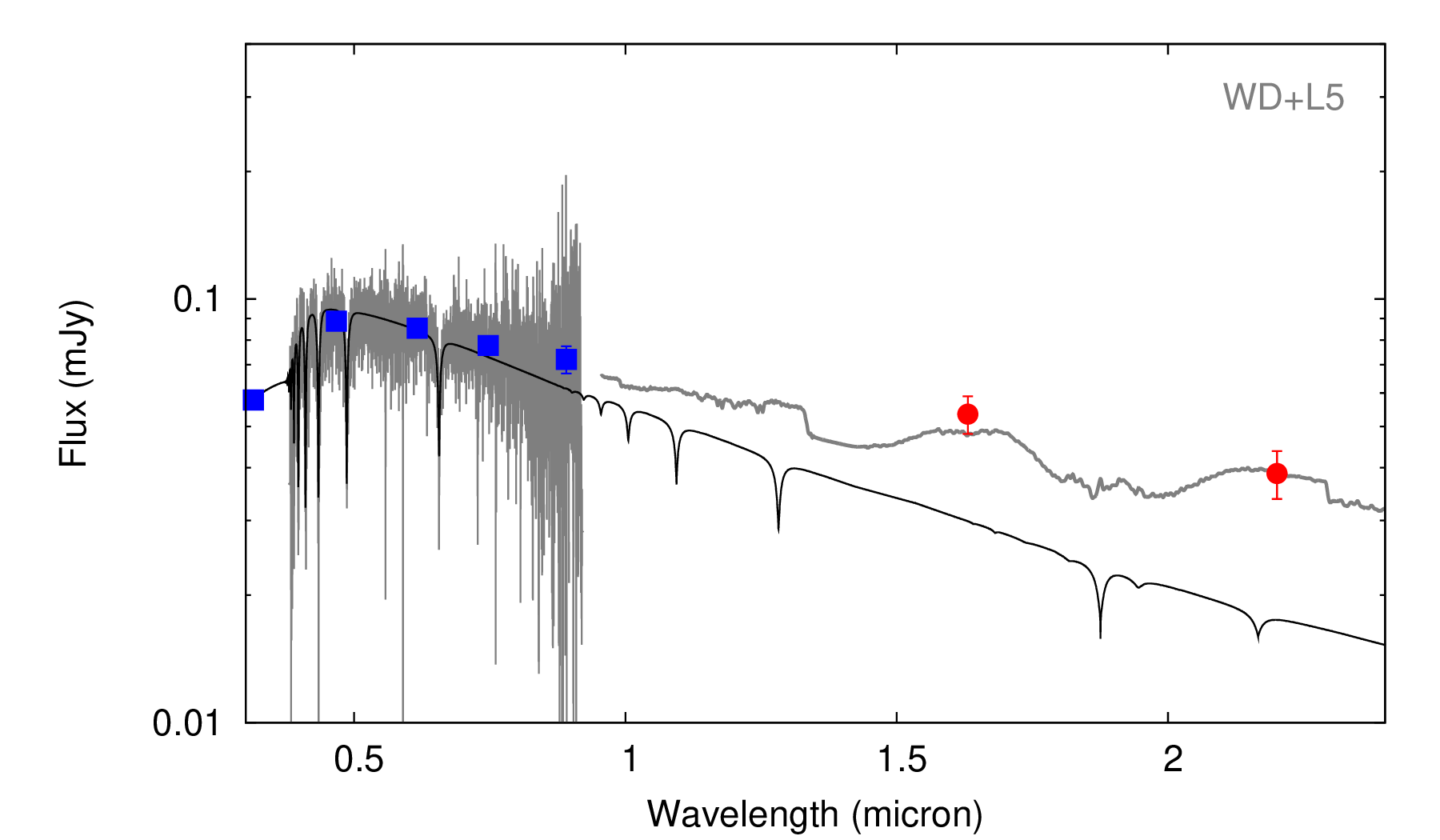}
  \caption{Two examples of DA WDs that show a NIR excess. SDSS photometry (squares), available UKIDSS photometry (circles), a model atmospheric spectrum (black solid), the SDSS spectrum (light grey) and a composite WD + companion model (as labeled) are plotted for each.}
  \label{103452}
\end{figure}

\subsection{Candidate selection}

Isolated WDs and candidate binaries were identified using the methods
described in Steele et al. (2011, in prep.). For UKIDSS DR8 the cross
correlation produced 3109 exclusive matches with the SDSS DR4
catalogue \citep{SDSSDR4}, 
and 163 with the McCook \& Sion catalogue (MS99, \citealt{McCook}).  

In brief, we have used Pierre Bergeron's synthetic colours and
evolutionary sequences for DA
WDs\footnote{\url{http://www.astro.umontreal.ca/~bergeron/CoolingModels/}}. The
model grid extends from $1500<T_{\rm eff}<100,000$\,K and
$7.0<$log\,$g<9.0$ but was linearly extrapolated to $6.5<$log\,$g<9.5$
to include the lower and higher mass WDs, but to preclude any possible
sub-dwarf contamination of the sample. This provided theoretical
absolute magnitudes for the SDSS $i'$ and the 2MASS $JHK_{\rm S}$
filters, as well as an estimated distance to the WD and cooling age of
the system. The 2MASS magnitudes were converted to the corresponding
UKIDSS filters using the colour transformations of \cite{carpenter01}. Out of the 2675 WDs classified by EIS06 and MS99 as DA WDs,
1040 had both $H$ and $K$-band photometry and fell within the extended
model grid.



We identified candidate stars as those showing at least a $>3\sigma$
excess in the UKIDSS $H$ and $K$-bands (or $K$-band only) when
compared to the model predicted values. 314 such excesses were
identified, with 275 of these accounted for as previously identified
binaries from an optical excess in SDSS (or other optical data). This
left 39 candidate WDs with an apparent NIR excess. Ten of these appear
to be caused by foreground or background contamination (e.g.\ due to a
nearby red star or galaxy), 25 by a previously unidentified companion
and 4 by a debris dust disk.

Empirical models for low mass sub-stellar objects were then added to a
WD synthetic spectrum and these composites were compared to UKIDSS
photometry to obtain an approximate spectral type for the putative
companion. Due to errors in temperature, surface gravity and therefore
distance, estimates of the companion spectral types are likely within
$\pm$1 spectral types of the best fitting composite WD $+$ companion
model spectrum. The results break down as follows; 1 dM3, 2 dM5, 3
dM6, 5 dM7, 2 dM8, 1 dM9, 2 dL0, 2 dL1, 1 dL4, 3 dL5, 1 dL7, 1 dL8 and
1 dT3.

\subsection{Low Mass Star or Brown Dwarf?}

In order to assess whether or not each putative companion is a low
mass MS star or a BD an estimate of the mass of the secondary has been
calculated. Firstly an age of the WD was calculated by the addition of
the predicted WD cooling age and MS lifetime of the progenitor
star. This was estimated using the initial-final mass relationship of
\cite{dobbie06}, which is valid for initial masses $>1.6$\msun
\citep{kalirai08}. An approximate MS lifetime can then be
calculated from the models of \cite{girardi00}. It should be noted
that for WDs where $M_{\rm WD}<0.5$\msun\, it is highly likely that
the star has evolved through mass transfer and for these stars an age
can not be calculated through this method. However, this can be seen
as further evidence for the existence of a secondary star. For these
stars, a lower limit on the mass of the secondary is estimated by
using the cooling age of the WD.

\begin{figure}
  \includegraphics[width=15cm]{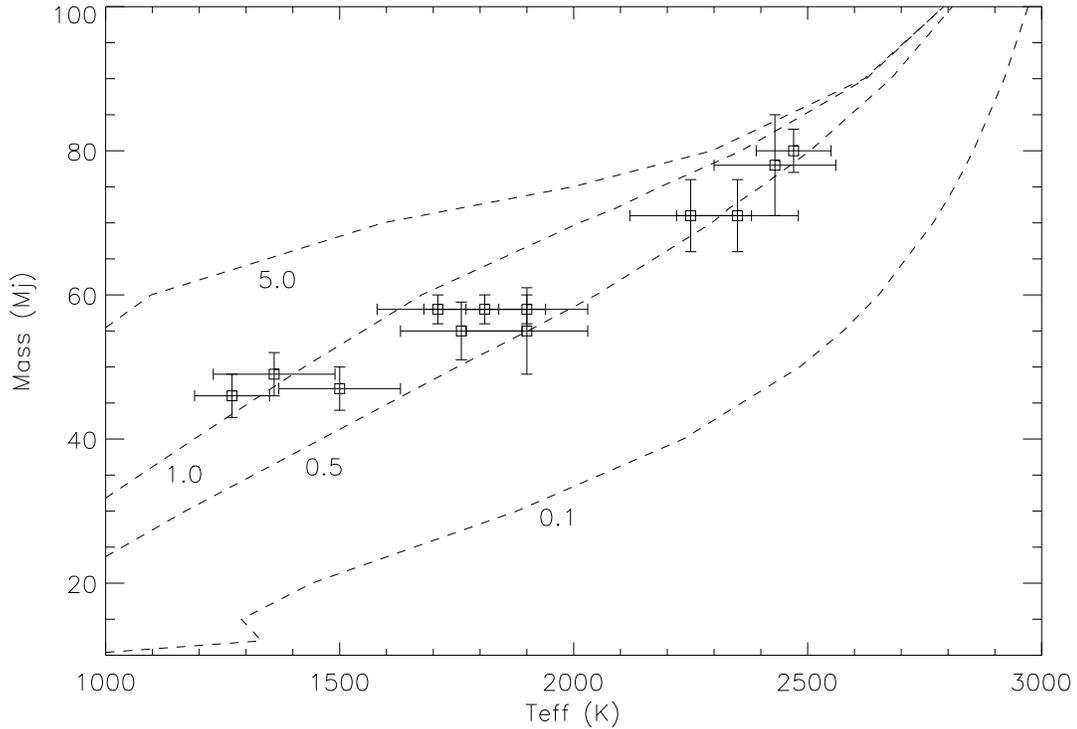}
  \caption{Predicted masses of companions below 80\mjup\, based on the Lyon Group models. The dashed lines indicate constant age in Gyr.}
  \label{mass_dusty}
\end{figure}

We then estimate a mass for the secondary by interpolating the Lyon
Group atmospheric models (e.g.~\citealt{dusty}),
given the age of the WD and an estimate of effective temperature of
the companion. We estimated these temperatures by comparison with
observed M, L and T dwarfs (\citealp{vrba}) and assuming an error of
$\pm1$ spectral type. The results are plotted in
Figure~\ref{mass_dusty}.

If we assume a value of 70\mjup\, as a conservative cut-off mass for
the stellar-substellar boundary, then 8-10 new DA WD$+$BD binaries
binaries have potentially been discovered (1 DA$+$L8 has been
spectroscopically confirmed; \citealt{steele09}).

\subsection{Limits on Unresolved Ultra-Cool Companions to DA WDs}

Atmospheric modeling successfully fitted 538 DA WDs to within
3$\sigma$ of both their UKIDSS $H$ and $K$-band photometry. We will
use these stars to estimate the limits on the spectral types of
unresolved companions to DA WDs and hence determine the sensitivity of
UKIDSS to both L and T dwarfs in WD binaries. The method used to
calculate these limits is detailed below.

In order to place a limit on the spectral type of the coolest
detectable unresolved companion to each DA WD we required the absolute
magnitudes for the spectral types of low mass objects ranging from
dL0-dT8. \cite{patten06} lists known M, L and T dwarfs with 2MASS
$JHK_{\rm S}$ photometry ranging from spectral types dM5-dT8 and also
has measured parallaxes for many of these.  We converted the 2MASS
$JHK_{\rm S}$ photometry for each object to the UKIDSS $JHK$
photometric system using the colour corrections of 
\cite{carpenter01}. Taking only those with a measured parallax, and thus an
estimated distance, the $K$-band photometry was scaled to 10\,pc for
each spectral type to obtain an estimated absolute $K$-band
magnitude. For multiple stars of the same spectral type we used an
average of the magnitudes. For the few spectral types that had no
parallax measurement we performed a linear interpolation to predict an
absolute magnitude.

\begin{figure}
  \includegraphics[height=7.5cm]{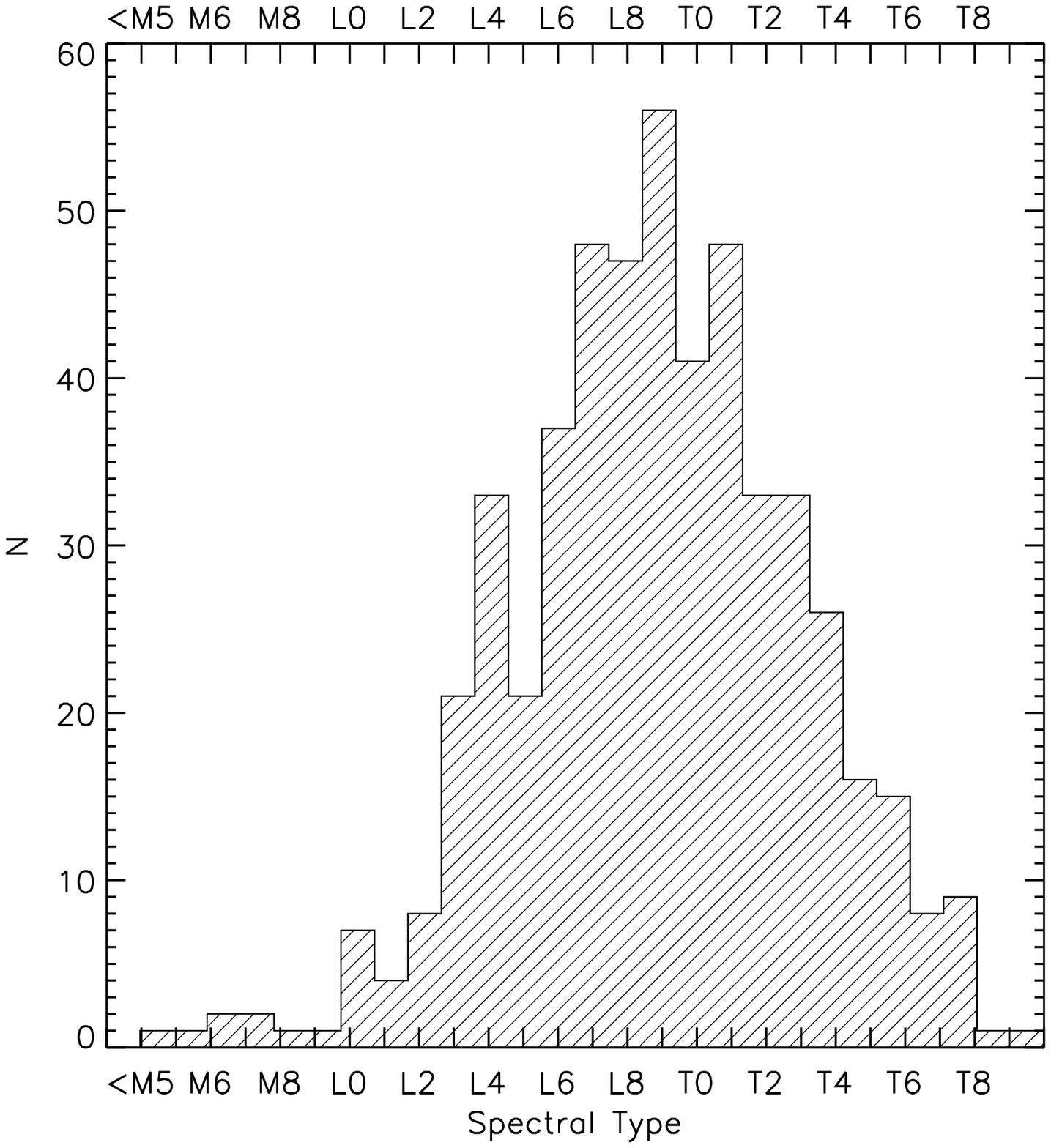}
  \includegraphics[height=7.5cm]{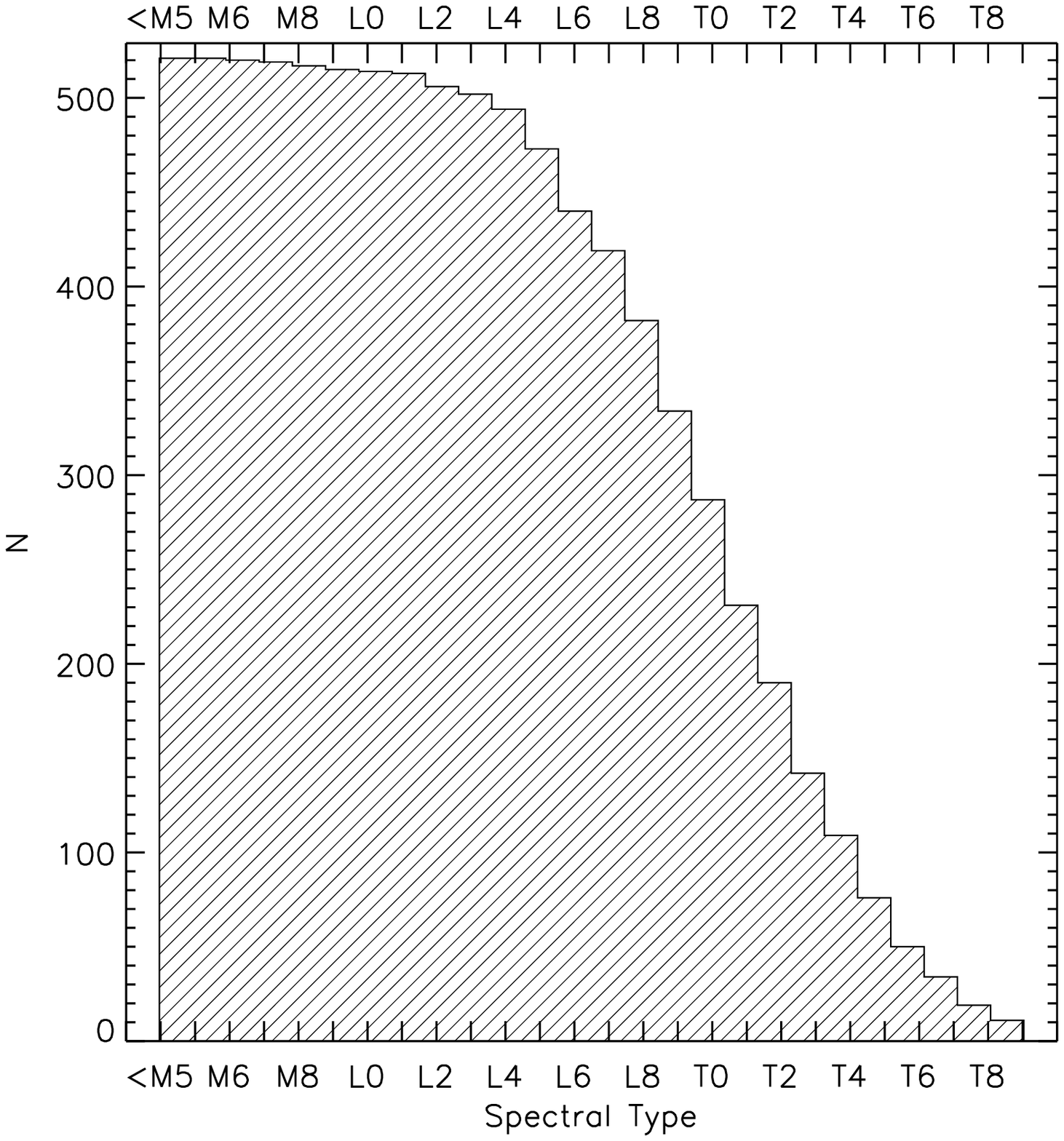}
  \caption{Left to right: (1) Distribution of limits on unresolved companions to the ``single'' DA white dwarfs (with well determined parameters) detected in UKIDSS DR8. (2) Same as (1) but plotted as a cumulative histogram.}
  \label{limits}
\end{figure}

For each apparently single DA WD we predicted the expected $K$-band
magnitude by folding the WD's atmospheric model through the UKIDSS
$K$-band filter transmission profile. The observed $K$-band magnitude
was not utilised, as any of the WDs could be in excess within the
observed errors. This would have resulted in an over-estimate of the
limiting spectral type for the companion to these stars.  We then
calculated a 3$\sigma$ detection limit for each WD by using the
observed errors listed in the UKIDSS DR8, and adding these to the
predicted magnitudes. The distance to each WD was calculated using the
Bergeron cooling models using the effective temperatures and surface
gravities produced from the automated fit of EIS06. For isolated DA
WDs only found in MS99, temperature estimates were obtained from the
literature or an automated fit to the stars' optical photometry. For
stars without a measured surface gravity, we assumed a
log\,$g=8.0$. This applied to only 11 WDs in the sample, which is not
a large enough number to create a distance bias that could later
affect the statistics. Subsequently, the grid of M, L and T dwarf
absolute magnitudes were scaled to the estimated distance of each WD
and added to the predicted WD $K$-band magnitudes until a match was
made with the 3$\sigma$ detection limit. The spectral types of all the
companion limits were summed and plotted as a histogram and a
cumulative histogram in Figure~\ref{limits}.

\subsection{Binary Statistics for WDs with Unresolved Substellar
  Companions}

In order to determine the fraction of WDs with BD companions, we must
first estimate the spectral type where the secondary becomes truly
substellar. This depends on the age of the secondary, which can be
estimated using the total age (MS lifetime $+$ cooling age) of the WD
primary. The average age of the WD sample is 1.8$\pm0.7$\,Gyr. After
$\approx0.1$\,Gyr BDs cool very slowly, so the average sample age can
be used to estimate a spectral type where the stellar/substellar
borderline occurs. Taking the upper limit on BD mass as 70\mjup, then
an upper limit for the BD spectral type can be taken as L4
(Figure~\ref{mass_dusty}).

In order to calculate the WD$+$BD binary fraction, we are going to
determine the effective number of detections. This value represents
the number of each spectral type detected if the survey were 100\%
efficient at detecting companions of all spectral types. The effective
number of detections can then be calculated by dividing the actual
number of detections by the sensitivity to each spectral type. For
example, if the survey sensitivity to companions of spectral types
earlier or equal to L0 was 50\%, and 1 L0 was detected, then the
effective number of detections is 2. This is then summed for all
spectral types to give the total effective number of BDs detected. The
BD companion fraction is then the effective number of detections
divided by the sample total. An error can be estimated by taking the
square root of the reciprocal number of detections. Thus, an upper
limit to the fraction of DA WDs with unresolved BD companions is
$f_{\rm WD+dL}\leq1.0\pm0.3$\%. A lower limit can be estimated by
assuming only 3 candidates (2 previously spectroscopically confirmed
and 1 proper motion candidate) are real. This gives a final range of
$0.3\leq f_{\rm WD+dL}\leq1.3$\%.

Although these statistics are not particularly robust, they are
suitable as a first approximation using the small numbers
available. In order to improve upon these numbers, the sample needs to
be enlarged. In the first instance, this will be done by UKIDSS, which
is set to be completed by 2012. Looking ahead, future infrared surveys
with VISTA and WISE will significantly 
add to the WD $+$ ultra-cool companion sample.

\begin{theacknowledgments}
PRS is supported by RoPACS, a Marie Curie Initial Training Network funded by the European Commission's Seventh Framework Programme. 
\end{theacknowledgments}



\bibliographystyle{aipproc}   


\IfFileExists{\jobname.bbl}{}
 {\typeout{}
  \typeout{******************************************}
  \typeout{** Please run "bibtex \jobname" to optain}
  \typeout{** the bibliography and then re-run LaTeX}
  \typeout{** twice to fix the references!}
  \typeout{******************************************}
  \typeout{}
 }

\end{document}